\documentstyle[aps]{revtex}

\def\deg{\ifmmode{^{\circ}}\else ${^{\circ}}$\fi}
\def\bi{\begin{itemize}}
\def\ei{\end{itemize}}
\def\ed{\end{document}}

\def\pri{^{\, \prime}}
\def\prii{^{\, \prime\prime}}
\def\cf#1{\ifmmode{\cal #1}\else${\cal #1}$\fi}

\def\be{\begin{equation}}
\def\ee{\end{equation}}
\def\beas{\begin{eqnarray*}}
\def\eea{\end{eqnarray}}
\def\bea{\begin{eqnarray}}
\def\eeas{\end{eqnarray*}}
\def\tfrac#1#2{{\textstyle\frac{#1}{#2}}}
\def\thalf{\tfrac{1}{2}}
\def\gev{\ifmmode{\mbox{GeV}}\else GeV\fi}
\def\es{{\rm erg\ s}^{-1}}

\def\evmss{\ \mbox{eV}^2\mbox{m}^{-2}\mbox{s}^{-1}\mbox{sr}^{-1}}
\def\lf{L_{41}}
\def\ton{T_{\rm on}}
\begin{document}

\twocolumn
\renewcommand{\topfraction}{1.0}
\twocolumn[\hsize\textwidth\columnwidth\hsize\csname
@twocolumnfalse\endcsname

\title{An Auger test of the Cen A model of highest energy cosmic rays}
\author{Luis A. Anchordoqui$^a$, Haim Goldberg$^a$, and Thomas J. Weiler$^b$}
\address{$^a$Department of
Physics, Northeastern University, Boston, MA 02115}
\address{$^b$ Department of Physics \& Astronomy, Vanderbilt University,
Nashville, TN 37235}

\maketitle

\begin{abstract}

If, as recently proposed by Farrar and Piran, Cen A is the source
of  cosmic rays detected above the Greisen-Zatsepin-Kuz'min
cutoff, neutrons are $\approx 140$ more probable than protons to
be observed along its line of sight. This is because the proton
flux is rendered nearly isotropic by ${\cal O}(\mu$G)
intergalactic magnetic fields. With the anticipated aperture of
the Southern Auger Observatory, one may expect on the order of 2
neutron events/year above $10^{20}$ eV in the line of sight of
Cen A.

\end{abstract}

\vskip2pc]

The energy spectrum of cosmic rays (CRs) is well fitted by power
laws with increasing index for energies above $4 \times 10^{15}$
eV (the ``knee'') flattening again above $5 \times 10^{18}$ eV
(the ``ankle''), yielding the overall shape of a leg. Over the
last third of the century, ingenious installations with large
effective areas and long exposure times---needed to overcome the
steep falling flux---have raised the tail of the spectrum up to
an energy of $3 \times 10^{20}$ eV, with no evidence that the
highest energy recorded thus far is Nature's upper limit
\cite{reviews}. The origin of these extraordinarily energetic
particles continues to present a major enigma to high energy
physics \cite{bs}.

The main problem posed by the detection of CRs of such energy (if
nucleons, gammas, and/or nuclei) is energy degradation through
inelastic collisions with the universal radiation fields
permeating the universe. Therefore, if the CR sources are all at
cosmological distances, the observed spectrum must virtually end
with the Greisen-Zatsepin-Kuz'min (GZK) cutoff at $E \approx 8
\times 10^{19}$ eV \cite{gzk}. The spectral cutoff is less sharp
for nearby sources (within 50 Mpc or so). The arrival directions
of the trans-GZK events are distributed widely over the sky, with
no plausible  counterparts (such as sources in the
Galactic Plane or in the Local Supercluster). Furthermore, the
data are consistent with an isotropic distribution of sources in
sharp constrast to the anisotropic distribution of light within
50 Mpc \cite{hillas}. The difficulties encountered by
conventional acceleration mechanisms in accelerating particles to
the highest observed energies have motivated suggestions
that the underlying production mechanism could be
of non-acceleration nature. Namely, charged and neutral
primaries, mainly light mesons (pions) together with a small
fraction (3\%) of nucleons, might be produced at extremely high
energy by decay of supermassive elementary $X$
particles ($m_X \sim 10^{22} - 10^{28}$ eV) \cite{top-down}.
However, if this were the case, the observed spectrum should be
dominated by gamma rays and neutrinos, in contrast to current
observation \cite{ave}! Alternative explanations involve
undiscovered neutral hadrons with masses above a few GeV
\cite{CFK}, neutrinos producing nucleons and photons via resonant
$Z$-production with the relic neutrino background \cite{z}, or
else neutrinos attaining cross sections in the millibarn range
above the electroweak scale \cite{nu}. A controversial correlation
between the arrival direction of CRs above $10^{20}$ eV
and high redshift compact radio quasars seems to support these
scenarios \cite{FB}.

Over the last few years, it has become evident that the observed
near-isotropy of arrival directions can be easily explained if
even the highest energy cosmic rays propagate diffusively,
camouflaging a unique source only a few Mpc away \cite{teapot}.
Within this framework, the particles experience large deflections
through randomly oriented patches of strong magnetic fields ${\cal
O}(\mu$G) \cite{muG,clustering}. Recently, Farrar and Piran (FP)
\cite{farrar-piran} noted that an extragalactic magnetic field of
$\sim 0 .3 \,\mu$G would bend CR paths sufficiently, allowing one
to trace back trans-GZK orbits from the Earth's
northern-hemisphere to the southern radio galaxy Cen A. Moreover,
they show that the flux of Cen A at $10^{19}$ eV (at Earth) is
comparable to that of all other sources in the universe, and
assuming a diffuse propagation of particles above this energy
they predict a CR anisotropy of order 7\% (or less). Both
estimates strongly support the single-source hypothesis. If this
is the case, and the absence of the GZK cutoff is a reflection of
our coincidental position near Cen A ($d \approx 3.4$ Mpc), it
must be that the emission of uncharged particles from Cen A
should render an enhancement of the CR flux in the southern
hemisphere.

Cen A is a complex Fanaroff-Riley (FR) I \cite{fr} radio-loud
source ($l \approx 310^\circ$, $b \approx 20^\circ$) identified
at optical frequencies with the galaxy NGC 5128 \cite{israel}.
The radio morphology is intricate with large non-thermal radio
lobes. In particular, the structure of the northern middle lobe
resembles the ``hot spots'' which exist at the extremities of
FR-II galaxies \cite{burns}, although for Cen A the brightness
contrast (hot spot to lobe) is not as extreme as in {\em e.g.}
Cyg A \cite{cyg}. The energetics of acceleration in hot spots were
discussed in \cite{rachen-b}. The criteria were applied in
\cite{cena} to show the plausibility of attaining trans-GZK
energies  in the hot spot of Cen A. Moreover, EGRET measurements
\cite{EGRET} of the gamma ray flux for energies $>100$ MeV allow
an estimate $L_{\gamma} \sim 10^{41}\ \es$ for the source
\cite{aharonian}. This value of $L_{\gamma}$ is consistent with
an earlier observation in the TeV-range during a period of
elevated activity \cite{g}, and is considerably smaller than the
estimated bolometric luminosity $L_{\rm bol}\sim
10^{43}\es$\cite{israel}.

CR ``lore'' convinces us that the TeV $\gamma$-ray emission is a
result of synchrotron radiation of electrons or protons of still
higher energy \cite{bst,pic}. Strictly speaking, the observed
$\gamma$-radiation is related to: (i) the development of pairs
cascades triggered by secondary photopion products that cool
instantaneously via synchrotron radiation (ii) the synchrotron
radiation of protons itself that becomes a very effective channel
to produce  high energy $\gamma$-rays above $10^{19}$ eV. There
are plausible physical arguments \cite{pic,waxman} as well as
some observational reasons \cite{steve-richard} to believe that
when proton acceleration is being limited by energy losses, the
CR luminosity $L_{\rm CR}\approx L_{\gamma}$. The low ratio
$L_{\gamma}/L_{\rm bol}$ thus leads us to assume that both ultra
high energy CR and $\gamma$ production take place in the lobes
with the bulk of the softer radiation coming from the core.

Following FP we introduce
$\epsilon$, the efficiency of ultra
high energy CR production compared to high energy $\gamma$ production---from
the above, we expect $\epsilon\simeq 1.$ Using equal power per
decade over the interval $1 \times 10^{19}{\rm eV}<E<4\times 10^{20}{\rm
eV}$, we estimate a source luminosity
\begin{equation}
\frac{E^2 \,dN^{p+n}_0}{dE\,dt} \, \approx 1.7\,\epsilon\lf \,10^{52}
{\rm eV/s} \label{1}
\end{equation}
where $\lf
\equiv$ luminosity of Cen A$/10^{41}\es$ and the subscript ``0'' refers
to quantities at the source.

Ignoring energy losses for the moment, the density of protons at the
present time $t$
of energy $E$
at a distance $r$ from Cen A (assumed to be continuously emitting at
a constant spectral rate $dN^{p+n}_0/dE\,dt$ from time $t_{\rm on}$ until the present)
is
\begin{eqnarray}
\frac{dn(r,t)}{dE} & = & \frac{dN^{p+n}_0}{dE\,dt}
\frac{1}{[4\pi D(E)]^{3/2}} \int_{t_{\rm on}}^t dt'\,
\frac{e^{-r^2/4D(t-t')}}{(t-t')^{3/2}}  \nonumber \\
 & = & \frac{dN^{p+n}_0}{dE\,dt}
\frac{1}{4\pi D(E)r} \,\,I(x),
\label{2}
\end{eqnarray}
where $D(E)$ stands for the diffusion coefficient,
$x = 4D\ton/r^2 \equiv \ton/\tau_D$, $\ton=t-t_{\rm on},$ and
\begin{equation}
I(x) = \frac{1}{\sqrt{\pi}} \int_{1/x}^\infty \frac{du}{\sqrt{u}} \,\, e^{-u}\ \ .
\end{equation}
In each ``scatter'', the diffusion coefficient describes an
independent angular deviation of particle trajectories whose
magnitude depends on the Larmor radius $R_{\rm L} = 100
E_{20}/B_{\mu{\rm G}}$ kpc, where $E_{20}=E/10^{20}{\rm eV}$,
$B_{\mu{\rm G}}= B/(1\ \mu{\rm G}).$ Extragalactic magnetic field
strengths and coherence lengths are not well established, but it
may be plausible to assume a Kolmogorov form for the turbulent
magnetic
field power spectrum with coherent directions on scales of 0.5 -
1 Mpc. One can then na\"{\i}vely estimate that protons with
energies $E < 10^{21} \ell_{\rm Mpc} B_{\mu{\rm G}}$ eV remain
trapped inside magnetic subdomains of size $\ell$, attaining efficient
diffusion when the wave number of the associated Alfv\'en wave is equal to
the gyroradius of the particle\cite{wentzel-drury}. With a
Kolmogorov spectrum this gives for a diffusion coefficient\cite{blasi-olinto}

\begin{equation}
D(E) \approx 0.048 \left( \frac{E_{20} \,\ell^2_{{\rm
Mpc}}}{B_{\mu{\rm G}}} \right)^{1/3} \,{\rm Mpc}^2/{\rm Myr}.
\end{equation}
Here, $\ell_{\rm Mpc}= \ell/(1\ {\rm Mpc}).$ For
$\ton\rightarrow \infty$, the density approaches its time-independent
equilibrium value $n_{\rm eq}$, while for $\ton= \tau_D=r^2/4D$,
$n/n_{\rm eq} = 0.16$.

A word about the validity of the diffusive approximation: one may easily
check that for $E=10^{19}$\ eV, $B=0.5\mu$G, $\ell = 0.5$ Mpc,
the diffusive distance traveled
$ c \tau_D=50$ Mpc $\gg d =3.4$ Mpc. For higher energies, the validity
of the diffusive
approach must be checked on a case-by case basis \cite{losses}. For these
purposes, in the case of a continuously emitting source, the definition of
a diffusion time is somewhat arbitrary. We will use $\tau_D,$
a choice consistent with simulations \cite{l2}.

To further constrain the parameters of the model, we  evaluate the
energy-weighted approximately isotropic proton flux at  $1.5\times 10^{19}$ eV,
which lies in the center of the flat
``low energy'' region \cite{farrar-piran} of the spectrum:
\begin{eqnarray}
E^3 J_p(E) & =
& \frac{Ec}{(4\pi)^2d\,D(E)} \frac{E^2 \,dN^{p+n}_0}{dE\,dt}\,
I(t/\tau_D)  \nonumber \\
 & \approx & 7.6 \times 10^{24} \, \epsilon \lf\, I
\,\, {\rm eV}^2 \, {\rm
m}^{-2} \, {\rm s}^{-1} \, {\rm sr}^{-1}. \label{jp}
\end{eqnarray}
In the second line of the equation, we have used the values of
$B\ \mbox{and}\ \ell$ as given in the previous paragraph. We
fix $\epsilon\, \lf\, I =0.40$, after comparing Eq.(\ref{jp}) to
the observed CR-flux: $E^3 J_{\rm obs}(E)  = 10^{24.5}$ eV$^2$
m$^{-2}$ s$^{-1}$ sr$^{-1}$ \cite{reviews}. With $\epsilon
\lf\simeq 1,$ this determines $I\simeq 0.40,$ and consequently
the required age of the source $\ton$ to be about 400
Myr~\cite{rmp}.  To maintain flux at the ``ankle'' for the same $\ton$,
we require an approximate doubling of $L_{\rm CR}$ at $5\times
10^{18}$ eV. Because of the larger diffusive time delay at this energy, this
translates into an increased luminosity in the early phase of Cen A.
The associated synchrotron photons are  emitted at energies $< 30$ MeV\cite{gs}.
The increase in radiation luminosity in this region is not inconsistent
with the  flattening of the spectrum observed  at lower energies\cite{ossecomtel}.

In current models of describing cosmic ray acceleration, the
principal mechanisms for energy loss are synchrotron radiation and
photopion processes \cite{bst,pic}. If the radiation energy
density of the source is sufficiently high, photopion production
leads to copious neutron flux (that can readily escape the
system) and associated degradation of the proton spectrum. This
occurs only  near the maximum proton energy\cite{bst}. It is
reasonable to assume that the ambient photon density of Cen A is
sufficiently high \cite{cena} so that near the end of the
spectrum the efficiency of neutron production $\epsilon_n$
becomes comparable to the proton channel $\epsilon_p$. We take
for granted that the proton spectrum cuts off at $4\times
10^{20}$ eV. Consequently, because of the leading particle effect
\cite{rachen-b}, we expect a cutoff in the neutron spectrum at
approximately $2\times 10^{20}$ eV. We adopt an energy of
$1\times 10^{20}$ eV as a lower cutoff on the neutron spectrum,
and simplify the discussion by assuming that in the narrow
interval $E_{20} \in [1,2]$ $\epsilon_n\approx \epsilon_p.$ The
neutron spectrum observed at Earth is further narrowed because of
decay {\em en route}. The decay length is $\lambda(E)=0.9\
E_{20}\, \mbox{Mpc}$ \cite{ppi}. Because of the exponential
depletion,  about 2\% of the neutrons survive the trip at
10$^{20}$ eV, and about 15\% at $2\times 10^{20}$ eV. We note at
this point that the increasing survival of neutrons at energies
above $1.5\times 10^{20}$ eV has as a consequence of the Cen A
model that the observed diffuse flux $E^3J_{\rm obs}(E)$ should
begin to decrease at these energies (unless other factors
contribute to an increase).

We may now estimate a signal-to-noise ratio for detection of
neutron CRs in the southern hemisphere, say at Auger\cite{auger}.
If we assume circular pixel sizes with  2\deg\ diameters, the
neutron events from Cen A will be collected in a pixel
representing a solid angle $\Delta \Omega({\rm Cen A}) \simeq
10^{-3}$ sr.
For Auger ($S = 3000$ km$^2$ detector with aperture 7000 km$^2$ sr
above $10^{19}$ eV), the event rate of (diffuse) protons coming
from the direction of Cen A (say in a $2^{\circ}$ angular cone)
is found to be
\begin{eqnarray}
\frac{dN_p}{dt} & = & S\, \Delta \Omega({\rm Cen A})\,\,
\int_{E_1}^{E_2} E^3 \, J_p(E) \, \frac{dE}{E^3} \nonumber \\
 & \approx & S\,  \Delta \Omega({\rm Cen A})\, \, <E^3 J_p(E)>\,
\frac{1}{2\,E_1^2} \nonumber \\
 & \alt & \frac{0.014}{E_{1,20}^2} \,\,\,\, {\rm events}/{\rm yr},
\label{ruido}
\end{eqnarray}
where we have assumed $E^3 J_p(E)$ to be (approximately) constant
up to at least
$E \approx 3 \times 10^{20}$ eV, in agreement with the observed
isotropic flux in this region, $E^3\,J_{\rm
obs}(E) = 10^{24.5 \pm 0.2} \evmss$
\cite{reviews}. The  neutron rate
\begin{eqnarray}
\frac{dN_n}{dt} & = & \frac{S }{4 \pi d^2} \,\,\int_{E_1}^{E_2}
\frac{dN_0^n}{dE dt} \, e^{-d/\lambda(E)} \nonumber \\
    & = & \frac{S }{4 \pi d^2} \,\,\int_{E_1}^{E_2}
    \frac{E^2\,dN_0^n}{dE\,dt} \, \frac{dE}{E^2}\, e^{-d/\lambda(E)} \nonumber \\
& = & 116 \, \epsilon_n \lf\int_{E_{1,20}}^{E_{2,20}}
\frac{dE_{20}}{E_{20}^2} \, e^{-d/\lambda(E)}\,\,\,\,  {\rm
events}/{\rm yr}, \label{7}
\end{eqnarray}
is potentially measurable. For $E_{20}\in[1,2]$ we expect
\begin{equation}
\frac{dN_n}{dt} \approx 4 \,\,\epsilon_n \lf \,\,\,\, {\rm events}/{\rm yr}
\label{pipi}
\end{equation}
arriving from the Cen
A direction of the sky. With
$\epsilon_n\lf\approx 1/2,$ this gives about 2 direct events per
year, against the negligible background of
Eq.(\ref{ruido})\cite{more}. Thus, in a few years running (of
Auger) the FP hypothesis of Cen A as the primary source of all
trans-GZK CRs {\em can be directly tested}.

We now address the question of anisotropy. This can be found by computing
the incoming current flux density $D\nabla n$ as viewed by an observer
on Earth, and one finds for a continuously-emitting source
a distribution $\sim(1+\alpha \cos(\theta))$
about the direction of the source at angle $\theta$ to the zenith, where
\begin{equation}
\alpha = \frac{2D(E)}{cr}\cdot \frac{I\pri}{I}.
\label{anisotropy}
\end{equation}
Here,
\begin{equation}
I\pri(x) = \frac{1}{\sqrt{\pi}} \int_{1/x}^\infty du\ \sqrt{u}
\,\, e^{-u}, \ee with $x=\ton/\tau_D$, and $I$ was defined in
Eq.(3)\cite{fpanis}. For our choices of $B$ and $\ell,$ $T_{\rm
on}=400$ Myr, we find for $E=10^{19}$ eV $(E=10^{20}$ eV) that
$\alpha = 0.04\ (\alpha = 0.07).$

It should also be remarked that the neutrons that are able to
decay will beget secondary proton diffusion fronts with asymmetry 
parameters given by  
\begin{equation}
\alpha = \frac{2D(E)}{cr}\cdot \frac{I\prii}{I}\  \ \ ,
\label{anisotropya}
\end{equation}
where
\begin{eqnarray}
I\prii(x) = \frac{1}{4\sqrt{\pi}\kappa} \int_{1/x}^\infty &\frac{du}{u^{3/2}}&
\left[  \left((1-\kappa)u + \thalf\right)\ e^{-(1-\kappa)^2u}\right.\nonumber\\
&-&\left.\left((1+\kappa)u + \thalf\right)\ e^{-(1+\kappa)^2u}\right]
\label{anisa}
\eea
and $\kappa=\lambda(E)/r,$ $\lambda(E)$ being the neutron decay length given
after Eq.(3).
In spite of the complicated nature of Eq.(\ref{anisa}), the results for
$\alpha$ are very
similar to the ones for the primary diffusion front given above.

All in all, the Southern Auger Observatory will be in a gifted
position to explore Cen A, providing in few years of operation
sufficient statistics to probe extragalactic magnetic fields
below the present observational upper limit ${\cal O} (\mu$G).
The potential detection of the neutrons at Auger can subsequently be
validated by the larger aperture EUSO and OWL orbiting detectors\cite{eusowl}.
Additionally, if FP's hypothesis is confirmed, it would
constitute a robust evidence that all FR radiogalaxies
produce extremely high energy CRs. Furthermore, our next-door
radiogalaxy could provide a profitable arena for particle physics.

In closing, we wish to comment briefly on some published CR observations relevant
to this work. A small excess
of flux at $10^{15}$ eV (detected at the Buckland Park field
station\cite{australia}) that reached the Earth preferentially from the direction
of Cen A could militate against FP's hypothesis.
At this energy the photon
flux will be completely damped through interactions with  the cosmic
microwave background \cite{gamma}.
Therefore, if CRs propagate diffusively one expects no deviation
from isotropy on the (extragalactic) CR spectrum (except for a
neutrino flux  peaked along the line of sight).
However, as far as we are aware, such anisotropy was not
confirmed by the Sydney University Giant Air Shower Recorder
(SUGAR) \cite{sugar}. Furthermore, the random arrival directions of the
southern highest energy CRs seem to back up the above-outlined
model.

\hfill

We would like to thank Pasquale Blasi and G\"unter Sigl for some stimulating
remarks concerning the diffuse propagation of cosmic rays.
This work was partially supported by CONICET (LAA), the National
Science Foundation (HG), and the U.S. Department of Energy (TJW)


\begin{thebibliography}{99}
\bibitem{losses} It is straightforward to see that for 3.4 Mpc the diffusion
time of any proton with energy above the photopion production
threshold is always less than the GZK-time. Thus, energy losses can be
safely neglected.
\bibitem{l2} G. Sigl, M. Lemoine and P. L. Biermann, Astropart.
Phys. {\bf 10}, 141 (1999). Comparing  with the simulation results for $r=10$
\ Mpc, $B=0.3\mu$G, $\lambda=1$\ Mpc, we obtain for $E=10^{19}$ eV that
$\tau_D=750$ Myr, in the midregion of the time-delay found in this reference.
For $E=10^{20}$\ eV (where energy losses are still small),  $\tau_D$=350 Myr,
not inconsistent with
the results of this reference, although at the margin of the distribution of
time-delays for this energy. This is indicative of  the transition to Bohm
diffusion.
\bibitem{ppi}The neutron beam suffers additional depletion because
of interaction with
the cosmic microwave background. The interaction length is approximately
6 Mpc, so that after 3.4 Mpc, 45\% of the neutrons interact. Half
go back into neutrons, so this 22\% depletion
effect falls  within errors.
\bibitem{more}An increase in the maximum energy attainable at the
source will shift the end of the spectrum to higher energy. This
will lead to significant enhancement of the neutron flux (at the
higher energy) because of a greater survival probability.
\end{thebibliography}
\end{document}